\documentclass[10pt,twocolumn]{article}
\usepackage{graphicx}
\usepackage{amsmath}

\title{\bf Coupled parametric processes in binary nonlinear  photonic structures}

\author{M.Yu. Saygin$^{1,2,\ast}$ and A.S. Chirkin$^{1}$\\
{\small {\it 	$^{1}$Faculty of Physics and International Laser Center, M.V. Lomonosov Moscow State University, Moscow, Russia } }   \\
{\small {\it	$^{2}$Quantum Radiophysics Department, P.N. Lebedev Physical Institute of Russian Academy of Sciences, Moscow, Russia } }\\
{\small {\it $^{\ast}$e-mail: saygin@physics.msu.ru } } \hspace{11.5cm}
}

\date{}

\begin{document}


\twocolumn[
  \begin{@twocolumnfalse}
    \maketitle
    \begin{abstract}
      We study parametric interactions in a new type of nonlinear photonic structures, which is realized in the vicinity of a pair of nonlinear crystals. In this kind of structure, which we call binary, multiple nonlinear optical processes can be implemented simultaneously, owing to multiple phase-matching conditions, fulfilled separately in the constituent crystals. The coupling between the nonlinear processes by means of modes sharing similar frequency is attained by the spatially-broadband nature of the parametric fields. 
We investigate the spatial properties of the fields generated in the binary structure constructed from periodically poled crystals for the two examples: 
1) single parametric down-conversion, and 
2) coupled parametric down-conversion and up-conversion processes. The efficacy of the fields’ generation in these examples is analyzed through comparison with the cases of traditional single periodically poled crystal and aperiodic photonic structure, respectively. 
It has been shown that the relative shift between the periodic crystal lattices has a crucial effect on the generated spatial field spectrum and the overall efficiency. In addition, the influence of the inter-crystal distance on these characteristics has been studied. 
Therefore, our study suggests that one can construct optical elements with sophisticated nonlinear properties from simpler elements without significant sacrifice of the efficacy.
    \end{abstract}
  \end{@twocolumnfalse}
]

\vspace{1cm}

\section{Introduction}\label{sec:introduction}

Nonlinear optical structures play an important role in modern science and technology, 
since it forms the basis for generation of optical fields with unique properties~\cite{Harder,HarrisHau,Zhang,Fulop} and provides convenient means to manipulate them~\cite{Hogstedt,Hall,Vasilyev,Guerreiro}.

Among a variety of nonlinear optical processes, the parametric down-conversion (PDC) process~\cite{Boyd} is extensively exploited today.
In particular,
different types of amplifiers~\cite{AkhmanovKhokhlov,Kingston,Giordmaine,AkhmanovKovrigin,Xu,Choi,Vaughan}  and 
wavelength-tunable oscillators~\cite{Harris,Eckardt,Stro,Reid} are based on the PDC implemented in media with quadratic susceptibility.
In the quantum domain, the PDC serves as a source of nonclassical states of light~\cite{Fedorov,ChuuHarris,Chuu}, in particular,  entangled states~\cite{Magnitskiy,Eberly} --- a  resource for  quantum-enhanced technologies; the squeezed quantum states, which are related to quadrature entanglement, are also produced by the PDC process~\cite{Andersen,Grote}. 
In addition to PDC, the sum-frequency generation (SFG) process is also implemented in the quadratic  media, thereby providing means to convert and  manipulate optical fields, which has been  in both classical~\cite{Hogstedt}  and quantum regimes~\cite{Vasilyev}.

Generally, the  phase-matching conditions imposes major limitation to the implementation of nonlinear interactions~\cite{Boyd}. 
The birefringent phase-matching which only relies on favorable material properties has very limited use due to material's constraints, whereas the quasi-phase-matching (QPM) technique~\cite{Armstrong,Phillips}, does not posses such limitations and is widely exploited today. 
Currently, the crystals with inversion of the susceptibility sign in one dimension are usually dealt with in applications~\cite{Armstrong}.  The lattice of periodically modulated crystals allows to compensate for single phase mismatch~\cite{Nikogosian}, or, at certain conditions, multiple mismatches~\cite{Grechin}. Chirped nonlinear crystals --- a kind of aperiodic structures --- are used for adiabatic generation and conversion of optical fields~\cite{Arbore,Harris2}. Extra control over the spatial and temporal spectrum of the generated fields is gained using 2D and 3D photonic structures~\cite{Phillips2,Saltiel,Vyunishev}.

In this work, we are interested in implementing simultaneously multiple processes in a single photonic structure, which is motivated by the necessity of monolithic multicolor optical sources that can be developed with these structures.  In the classical regime, mutually coupled PDC and SFG processes produce optical fields that fall into distinctly located spectral regions, which is of applied interest an example of which is holography~\cite{Patents}. From the quantum perspective, the multicolor fields produced by these processes can carry sophisticated types of entangled states --- the building blocks of quantum-enhanced algorithms~\cite{Rodionov,ChirkinSaigin,ChirkinSaigin2}. 
In turn, these fields can be utilized in experiments where one needs an interface between distinct physical systems, for instance, to bridge the transition wavelengths of trapped atoms, that are used to perform quantum logic operation or quantum memory, and the telecommunication wavelengths to conveniently communicate with a distant party~\cite{Shahriar}.

Previously, the authors suggested one-dimensional structures as means to generate multicolor fields~\cite{ShutovChirkin,Chirkin,ChirkinShutov2}. 
In these aperiodic structures, the desired crystal lattice is constructed by 
superposing several lattice harmonics, each of which compensates single phase mismatch~\cite{ShutovChirkin,ChirkinShutov2}. 
Moreover, the inverse problem of finding the peak amplitudes of the lattice spectrum, that define the effective nonlinearity coefficients, has been solved analytically~\cite{ChirkinShutov2}.

Here, we investigate the alternative type of inhomogeneous structures that also enables implementation of multiple processes simultaneously. 
In contrast to the aperiodic nonlinear structures studied before, we suggest to implement coupled interactions in 2D structures, constructed from a pair of closely spaced crystals with generally different lattices, so that the resulted block of crystals form a binary structure.  
Owing to the multimodity of the field spatial spectrum, stemmed from the loosely constraints of phase-matching, coupling between nonlinear processes evolving in the neighboring crystals can be achieved. 
The difference of the coupling mechanism utilized in this study from the waveguide structures, where the localized eigenmodes are coupled evanescently has to be noted 
(see examples~\cite{Rai,Kruse} and references therein). 
Additionally, the capability to operate by large bandwidth fields makes the inhomogeneous crystal structures advantageous over waveguides, which spatial spectrum is limited by the eigenmodes profiles. 

We study the field generation in the proposed binary structures by analyzing nonlinear interactions comprising both the PDC and SFG processes, and compare the obtained results with the ones for the 1D nonlinear crystals.

The paper is organized as follows. In Section~\ref{sec:binary} we describe the binary structure under study. In Section~\ref{sec:degeneratePDC} single PDC process taking place in the binary structure constructed from crystals with equal but shifted lattices is investigated. 
Here, in Section~\ref{sec:vacuum}, we describe the link between the classical and quantum equations to account for vacuum fluctuations that seed the spontaneous PDC. Following that, in Section~\ref{sec:coupledprocesses}, we study the coupled processes of PDC and SFG, supported by the crystals individually. The conclusion is given in Section~\ref{sec:conclusion}.

\section{Binary nonlinear structure}\label{sec:binary}

\begin{figure}[htbp]
\centering
\includegraphics[width=0.9\linewidth]{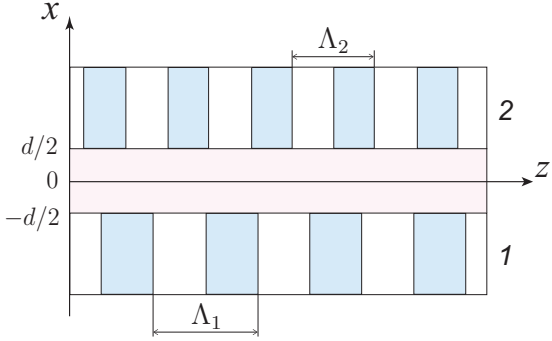}
\caption{Schematic of the binary nonlinear structure comprised of two periodically poled crystals with periods $\Lambda_1$ and $\Lambda_2$
separated by distance $d$. 
The transverse and longitudinal dimensions are marked by axes $x$ and $z$, respectively; 
the other transverse dimension, corresponding to $y$ axis (not shown), is irrelevant under the adopted approximations.}
\label{fig:fig1}
\end{figure}

The binary structure under study is illustrated in Fig.~\ref{fig:fig1}. 
The structure consists of two crystals (crystal 1 and crystal 2) placed at a distance $d$  (Fig.~1). We assume that the crystals' dimensions along axes $x$ and $y$ are large enough, thereby the peripheral effects are negligible. 
To fulfil the QPM conditions, the sign of the nonlinear susceptibilities of the crystals is modulated along the direction $z$, 
so that the corresponding quadratic susceptibilities read:  $\chi^{(2)}(z)=\chi_mg_m(z)$, where  $g_m(z)$ is the modulation function taking values <<-1>> and <<+1>>,   $\chi_m$ is the effective quadratic susceptibility, dependent on the material and the wavelengths involved in the processes,  $m$  is the index that mark the crystal ($m=1,2$).

Given the structure properties along the transverse direction (axis x) are defined by the  crystals and the medium in between them, we have the following expression for the spatial-dependent effective quadratic susceptibility:
	\begin{equation}\label{eqn:2Dnonlinearity}
		\begin{split}
			\chi(x,z)&=\theta\left(-x-\frac{d}{2}\right)\chi_1(z)+\theta\left(x-\frac{d}{2}\right)\chi_2(z)=\\
			=&\chi_1\theta\left(-x-\frac{d}{2}\right)g_1(z)+\chi_2\theta\left(x-\frac{d}{2}\right)g_2(z),
		\end{split}
	\end{equation}
where $\theta(x)$ is the Heaviside function. 
Notice, that at 
$\chi_1=\chi_2=\chi$, $g_1(z)=g_2(z)=g(z)$ and $d=0$ 
expression \eqref{eqn:2Dnonlinearity} turns into $\chi{}g(z)$ --- as for the case of traditional $1$D inhomogeneous crystal.

The linear optical properties of the binary structure are defined by the ones of the crystals and the medium in between them. In a general case when the constituent materials of the structure are different, both radiative and waveguide modes have to be considered, for rigorous analysis. Here, we assume the linear properties of the constituents identical, which can be the case of similar materials with the QPM conditions unfulfilled for the inter-crystal space. Thus, henceforth we take  $\chi_1=\chi_2=\chi$. 
These assumptions allow us to consider the task solely through the basis of plane waves.

We restrict our analysis to the periodically poled crystals, so that the modulations functions can be represented as 
	\begin{equation}
		g_m(z)=\text{sign}\left[\cos\left(\frac{2\pi}{\Lambda_m}z+\varphi_m\right)\right],
	\end{equation}
where $\Lambda_m$ is the lattice period of crystal $m$, 
$\varphi_m$ is the phase parameter, describing the longitudinal shift of crystal $m$ with respect to the position $z=0$, $\text{sign}[\cdot]$ is the signum function.

\section{Parametric down-conversion in the binary structure}\label{sec:degeneratePDC}

Let us consider a binary structure with both crystals supporting a single degenerate PDC process that obeys  the frequency relation:
	\begin{equation}\label{eqn:PDCfrequencies}
		\omega_p=2\omega_1,
	\end{equation}
where $\omega_p$ and $\omega_1$ are the pump and signal frequencies, respectively. 
Given the phase mismatch along the axis $z$:
$\Delta{}k=k_p-2k_1$, with $k_p$ and $k_1$ being the wavenumbers of the pump and signal,
the period of the compensating lattices is calculated: $\Lambda_1=\Lambda_2=\Lambda=2\pi/|\Delta{}k|$.
Since the lattices can be shifted with respect to each other in direction $z$,
we introduce the shift $\delta{}z$ that is related to the phase parameters $\varphi_m$:
$\delta{}z=\Lambda(\varphi_1-\varphi_2)/2\pi$. 
Therefore, the modulation functions of the lattices can be written as 
$g_1(z)=g(z-\delta{}z/2)$ and $g_2(z)=g(z+\delta{}z/2)$

To investigate the nonlinear dynamics of the PDC in the 2D structure, we adopt the slowly-varying amplitude approximation~\cite{Boyd} and  assume that the pump field is undepleted. 
Using \eqref{eqn:2Dnonlinearity} and taking diffraction into account, the equation reads:
	\begin{equation}\label{eqn:basic3Dequations}
		\begin{split}
			\frac{\partial{}A_1}{\partial{}z}+\frac{i}{2k_1}\Delta_{\perp}A_1=\\
			=\beta_dA_pA_1^{*}e^{-i\Delta{}kz}\left[\theta\left(-x-\frac{d}{2}\right)g\left(z-\frac{\delta{}z}{2}\right)\right.&\\
			+\left.\theta\left(x-\frac{d}{2}\right)g\left(z+\frac{\delta{}z}{2}\right)\right]&,
		\end{split}
	\end{equation}
where $A_1=A_1(\vec{\rho},z)$, $A_p=A_p(\vec{\rho},z)$ are the signal and pump field amplitudes, respectively, $\beta_d=\omega_1\chi/2n$ is the nonlinearity strength with  $\chi$
being the effective quadratic susceptibility of the crystals, 
$n$ is the refractive index at the signal frequency, 
$\Delta_{\perp}=\partial^2/\partial{}x^2+\partial^2/\partial{}y^2$ is the transverse Laplacian.

To simplify analysis, we use some convenient approximations. 
Firstly, even though \eqref{eqn:basic3Dequations} is three-dimensional, the binary structure is two-dimensional, thus, we shall solve the problem in two dimensions implying    
$\partial^2/\partial{}y^2=0$. 
Secondly, we take the pump wave to be plain, so that in equations that follow it is a constant value.

Thirdly, using the QPM conditions hold, the right-hand side of \eqref{eqn:basic3Dequations} can be simplified to eliminate the coordinate-dependent modulation functions. For this, the periodic function $g(z)$  is expanded into the discrete series 
(here, we approximate the finite length function $g(z)$ by  sum rather than  integral expansion):
	\begin{equation}\label{eqn:latticespectrum}
		g(z)=\sum_{m=-\infty}^{+\infty}g_m\exp(imq_zz),\qquad{}q_z=\frac{2\pi}{\Lambda},
	\end{equation}
where $g_m=2\sin(mD)/\pi{}m$ is the lattice spectrum amplitudes ($m\ne0$),
$m$ is the QPM order, $D$ is the duty cycle.
Further, we consider crystal lattices with $D=1/2$, so that $g_m=2/\pi{}m$.

Using \eqref{eqn:latticespectrum}, we obtain
	\begin{equation}\label{eqn:latticeexpansion}
		g(z\pm\delta{}z/2)e^{-i\Delta{}kz}=\sum_{m}g_m\exp\left[i(mq_z-\Delta{}k)z\pm{}imq_z\delta{}z/2\right]
	\end{equation}

In the case of   $q_z=\Delta{}k$, the dominant term in expansion~\eqref{eqn:latticeexpansion} is $g_1$.
Neglecting the remaining terms, which is equivalent to averaging over the structure length $L$,  the 
replacement  in~\eqref{eqn:basic3Dequations} follows:
	\begin{equation}\label{eqn:latticesubstitution}
		g(z\pm\delta{}z/2)e^{-i\Delta{}kz}\longrightarrow\overline{g}\exp(\pm{}i\psi),
	\end{equation}
where $\overline{g}=2/\pi$, $\psi=\Delta{}k\delta{}z/2=(\varphi_2-\varphi_1)/2$.

To analyze the field dynamics, it is convenient to work in the spatial spectrum domain. 
Using the Fourier expansion 
	\begin{equation}\label{eqn:FourierExpansion}
		A(x,z)=\int_{-\infty}^{+\infty}a(q,z)\exp(-iqx)dq,
	\end{equation}
\eqref{eqn:basic3Dequations} is rewritten as
	\begin{equation}\label{eqn:fourierPDCequation}
		\begin{split}
			&\frac{\partial{}a_1(q,z)}{\partial{}z}-i\frac{q^2}{2k_s}a_1(q,z)\\
			&=\sigma{}e^{i\varphi_p}\overline{g}\int_{-\infty}^{+\infty}\left(\tilde{\theta}_{-}(\kappa)e^{-i\psi}+\tilde{\theta}_{+}(\kappa)e^{i\psi}\right)a_1^{*}(\kappa-q,z)d\kappa,
		\end{split}
	\end{equation}
where $\sigma=\beta_d|A_p|$ is the coupling strength, $\varphi_p$ is the pump wave phase ($A_p=|A_p|\exp(i\varphi_p)$). 
In \eqref{eqn:fourierPDCequation} we have introduced the functions 
describing nonlinear interaction proceeding in the crystals $1$ and $2$:
	\begin{equation}\label{eqn:fouriertheta}
		\begin{split}
			\tilde{\theta}_{\pm}(\kappa)=\frac{1}{2\pi}\int_{-\infty}^{+\infty}\theta(\pm{}x-d/2)\exp(i\kappa{}x)dx&\\
			=\frac{1}{2}\left(\delta(\kappa)\pm\frac{i}{\pi\kappa}\right)\exp(\pm{}i\kappa{}d/2),&
		\end{split}
	\end{equation}
denoted by <<->> and <<+>> subscripts, respectively.

Substituting  \eqref{eqn:fouriertheta}  into  \eqref{eqn:fourierPDCequation}, we arrive at the following equation, describing the parametric interaction in the plane waves basis: 
	\begin{equation}\label{eqn:fourierPDC}
		\begin{split}
			\frac{\partial{}a_1(q,z)}{\partial{}z}-i\frac{q^2}{2k_1}a_1(q,z)&\\
			=\sigma{}e^{i\varphi_p}\overline{g}\left[\cos\psi{}a_1^{*}(-q,z)-\right.&\\
			-\left.\frac{1}{\pi}\int_{-\infty}^{+\infty}\sin\left(\psi+\frac{(q-q')d}{2}\right)\frac{a_1^{*}(-q,z)}{q-q'}dq'\right],&
		\end{split}
	\end{equation}
Without loss of generality, we set  $\varphi_p=0$, as  the field power grows more efficiently in this case.

It is instructive to consider some limiting cases of \eqref{eqn:fourierPDC}. 
In the first case, when the crystal interfaces are in contact, i.e.  $d=0$, the right-hand side of \eqref{eqn:fourierPDC} takes a simpler form:   $\sigma\overline{g}\left[\cos\psi{}a_1^{*}(-q,z)-\sin\psi{}H(a_1^{*}(-q,z))\right]$,
where  $H(f(q))=\frac{1}{\pi}\int_{-\infty}^{+\infty}\frac{f(q')}{q-q'}dq'$  is the Hilbert transform~\cite{King}. 
In addition to this, when the lattices are not shifted ($\psi=0$), we obtain the equation corresponding to the traditional case of periodically poled crystal~\cite{SayginChirkin,Kolobov}. 
In the second case, at arbitrary separation  $d$ and no shift ($\psi=0$), the right-hand side of  \eqref{eqn:fourierPDC} turns into  $\sigma\overline{g}\left[a_1^{*}(-q,z)-\frac{d}{2\pi}\int_{-\infty}^{+\infty}\text{sinc}((q-q')d/2)a_1^{*}(-q',z)dq'\right]$,   with the second term having the meaning of cutting out nonlinear interaction in the space between the crystals.

For the sake of comparison, we use the solution for the traditional case of single crystal with $d=0$ and $\psi=0$ as a benchmark.
The equation describing this  case is derived by setting $\psi=0$  and $d=0$, which yields the right-hand side of \eqref{eqn:fourierPDC}:  $\sigma\overline{g}a_1^{*}(-q,z)$. 
Also, the solution is compared with the one for the step-like structure, that implement nonlinear interactions in a half-space. 
$x\le{}0$-half-space is considered here, thus, 
taking  $\tilde{\theta}_{+}(\kappa)=0$  in \eqref{eqn:fourierPDCequation} yields the equation for the structure:
	\begin{equation}\label{eqn:PDChalfcrystal}
		\begin{split}
			\frac{\partial{}a_1(q,z)}{\partial{}z}-i\frac{q^2}{2k_1}a_1(q,z)&\\
			=\frac{1}{2}\sigma\overline{g}\left[a_1^{*}(-q,z)-iH(a_1^{*}(-q,z))\right].&
		\end{split}
	\end{equation}

Notice the factor of $1/2$ that occurs in the right-hand side of \eqref{eqn:PDChalfcrystal}.

\subsection{Solution method of  classical equations with vacuum input}\label{sec:vacuum}

For the analysis that follows, one needs to solve \eqref{eqn:fourierPDC} and \eqref{eqn:PDChalfcrystal}. 
We use the known form of solution for PDC evolving in the crystal without transverse modulation (\eqref{eqn:fourierPDC} at $d=0$, $\psi=0$)~\cite{SayginChirkin,Kolobov}:
	\begin{equation}\label{eqn:PDCsolutionform}
		a(q,z)=U(q,z)a_0(q)+V(q,z)a_0^{*}(-q),
	\end{equation}
where $a_0(q)=a(q,0)$ and  $a_0^{*}(-q)=a^{*}(-q,0)$ and   are the spectrum amplitudes at the crystal input,  $U(q,z)$ and $V(q,z)$  are the functions describing the solution. 
The explicit expressions in this case are also known~\cite{SayginChirkin,Kolobov}:
	\begin{equation}\label{eqn:standardPDCsolution}
		U(q,z)=\cosh\Gamma{}z+i\frac{\varepsilon}{\Gamma}\sinh\Gamma{}z,\quad{}V(q,z)=\frac{\sigma}{\Gamma}\sinh\Gamma{}z,
	\end{equation}
where $\varepsilon=q^2/2k_1$, $\Gamma=\sqrt{|\sigma|^2-\varepsilon^2}$.

The input-output form of solution \eqref{eqn:PDCsolutionform} is applicable for any input states, as far as the undepleted and plain pump wave assumpstion holds.
However, the quantum nature of the vacuum states, which is common in experiments and is implied in this paper, hinders obtaining solution directly. In this case, to calculate useful field characteristics the amplitudes  $a$ and $a^{*}$    have to be treated as quantum operators that obey appropriate commutation relations. In particular, from the commutation relations one can obtain the link between the functions  $U(q,z)$ and $V(q,z)$:
	\begin{equation}\label{eqn:PDCfromcommutation}
		|U(q,z)|^2-|V(q,z)|^2=1.
	\end{equation}
For example, ensure that the solution \eqref{eqn:standardPDCsolution} obeys \eqref{eqn:PDCfromcommutation}.

\begin{figure}[htbp]
\centering
\includegraphics[width=0.65\linewidth]{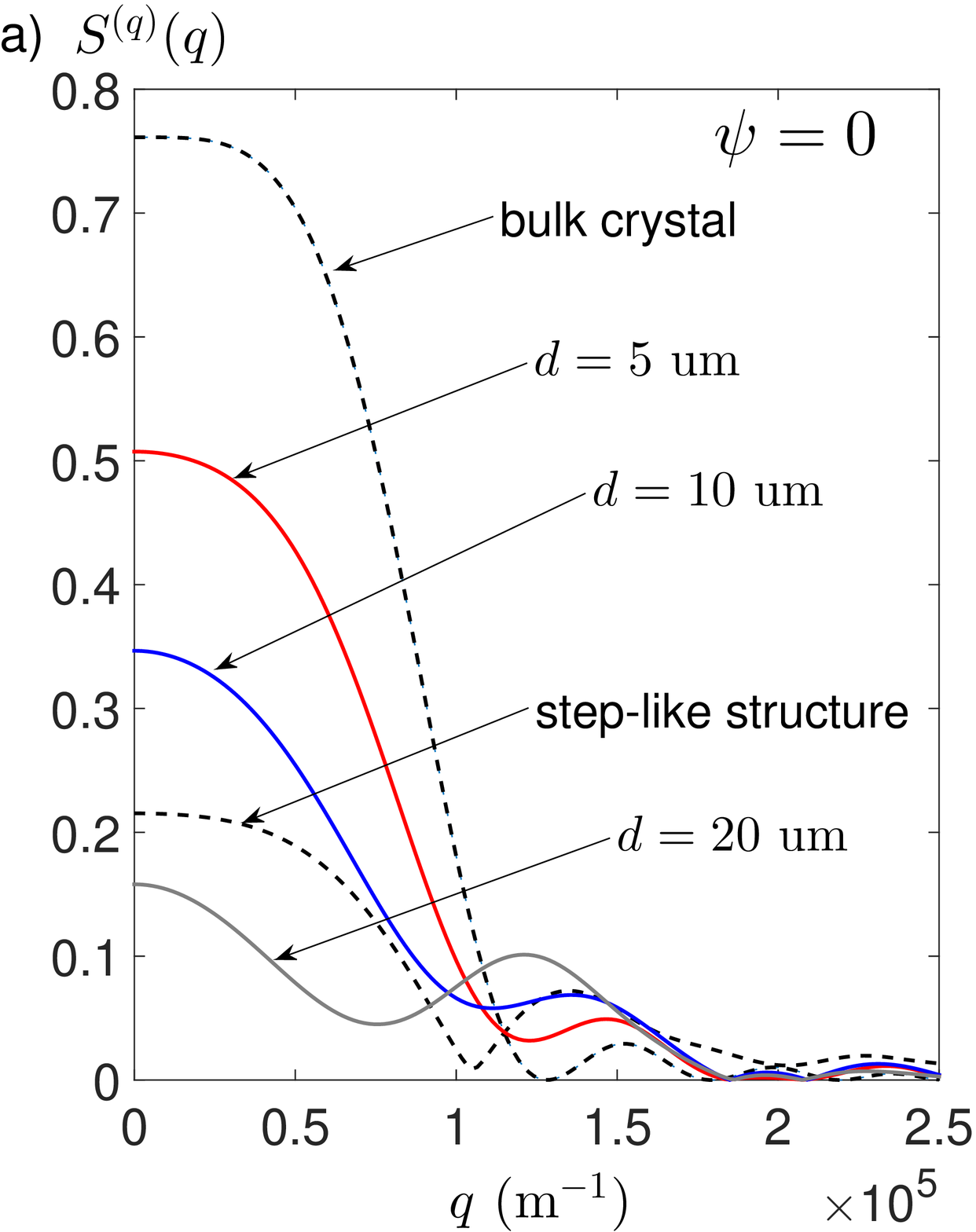}\\
\includegraphics[width=0.65\linewidth]{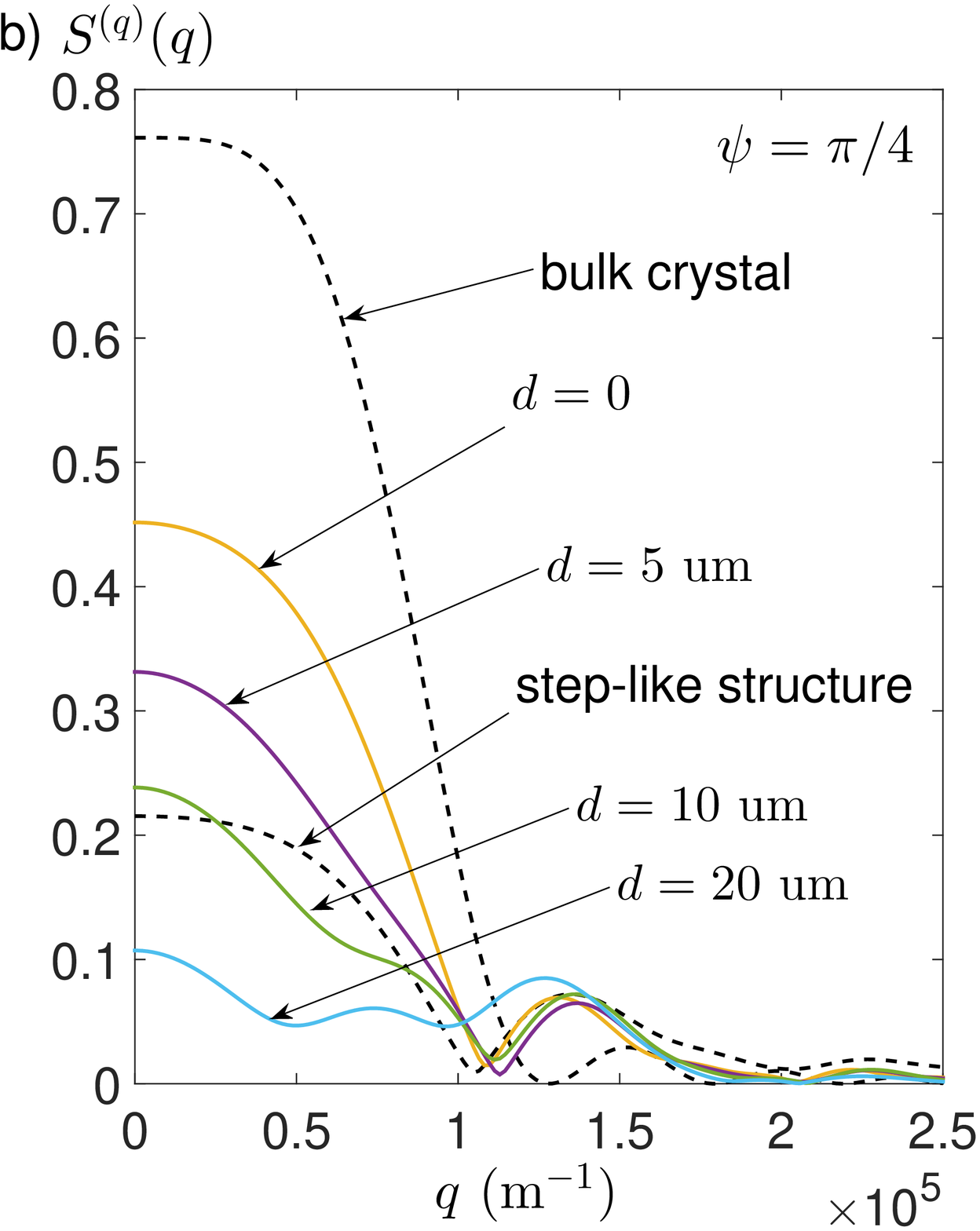}\\
\includegraphics[width=0.65\linewidth]{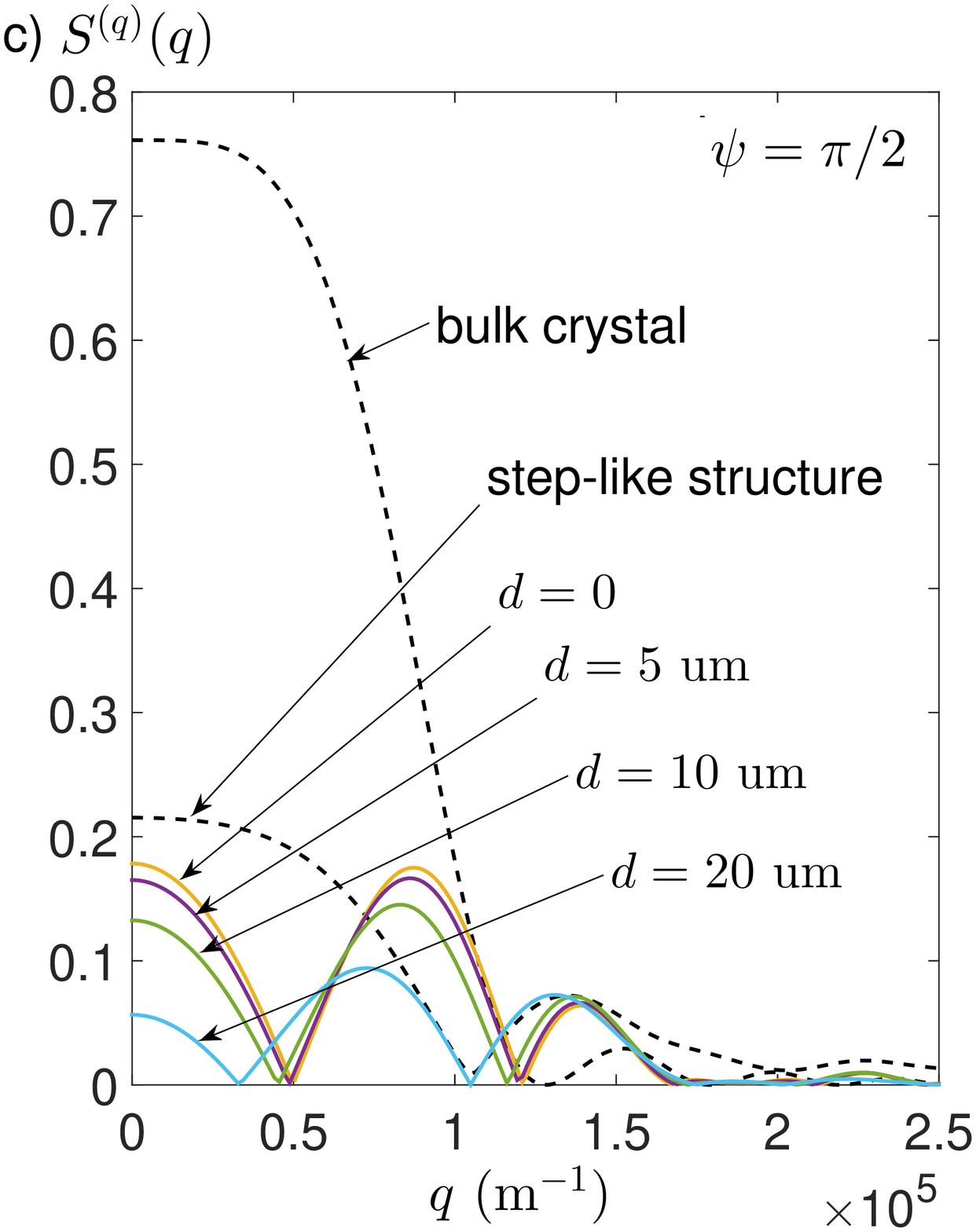}
\caption{The spectral density of the field generated in PDC in the binary photonic structure as a function of spatial frequency at a)  $\psi=0$ (without shift), b) $\psi=\pi/4$, c)  $\psi=\pi/2$ (maximum shift --- the domains of different signs are against each other). The dashed curves are the dependencies corresponding to PDC in standard crystal without transverse modulation and PDC taking place in the half-space. The graphics are plotted for the following parameters:  $\sigma=250$ m$^{-1}$, $\lambda_p=0.532$ um, $\lambda_1=2\lambda_p=1.064$ um, $L=0.5$ cm; the coupling coefficient value can be obtained, for example, in the lithium-niobate at $1$ kW pump power focusing in the  $100$ um diameter spot size~\cite{Boyd}.}
\label{fig:fig2}
\end{figure}

As for the field evolution in the binary structure, the solution of  \eqref{eqn:PDCsolutionform} can not be written explicitly similarly to~\eqref{eqn:standardPDCsolution}. 
Thus, we solved them numerically accounting for the caveat of quantum vacuum fluctuations at the input. 
Before using a numerical algorithm, we establish the relation between the fields in classical and quantum treatment.

Considering the input spatial amplitudes  $a_0(q)$ as statistically broadband and uniformly distributed, we obtain the mean $\langle{}a_0(q)\rangle=0$  for all $q$ and the correlations:
	\begin{equation}\label{eqn:PDCcorrelator}
		\langle{}a_0^{*}(-q')a_0(q'')\rangle=S_0(q')\delta(q'+q''),
	\end{equation}
where $S_0(q)=S_0$   is the spatial spectral density of the field.

Using \eqref{eqn:PDCfromcommutation} and \eqref{eqn:PDCcorrelator}, the spectral density at the output ($z=L$)  reads
	\begin{equation}
		S^{(c)}(q)=S_0(1+|V(q,z=L)|^2).
	\end{equation}
From the quantum point of view, however, replacing the amplitudes $a_0(q)$  and $a_0^{*}(q)$  with the corresponding operators    $\hat{a}_0(q)$  and $\hat{a}_0^{\ast}(q)$   yields different result~\cite{Kolobov}: $S^{(q)}(q)=S_0|V(q,z=L)|^2$. Given the spectral density obtained classically and quantum-mechanically, one can derive the following relation:
	\begin{equation}
		\frac{S^{(c)}(q)-S_0}{2S_0}=|V(q,z=L)|^2=S^{(q)}(q).
	\end{equation}
Therefore, we have shown how to calculate the spectral density of the field in PDC in a correct way, based on the characteristic obtained formally in the classical input assumption.

To solve \eqref{eqn:fourierPDC} numerically, 
statistical modeling of the vacuum by classical random noise with the uniformly distributed phase can be performed. 
However, in our case, this approach is inefficient, since it requires numerical propagation of \eqref{eqn:fourierPDC} for the series of randomly generated input amplitude values and the final result  is significantly noisy due to finite sampling.
Fortunately, the linear form of \eqref{eqn:PDCsolutionform} enables a more efficient
approach. For this, assuming constant values of  amplitudes $a_0(q)=a_0$  we construct the following combination:
	\begin{equation}\label{eqn:PDCfprmalinterference}
		\begin{split}
			S^{(det)}(q;a_0)=a^{*}(-q,z)a(q,z)&\\
			=|a_0|^2\left(1+2|V(q,z)|^2\right)&\\
			+U(q,z)V^{*}(q,z)a_0^2(q)+U^{*}(q,z)V(q,z)a_0^{*2}(-q),&
		\end{split}
	\end{equation}
which is calculated by single numerical propagation of \eqref{eqn:fourierPDC}. 
At random input values, the last two terms in the right-hand side of \eqref{eqn:PDCfprmalinterference} are vanished after statistical averaging. 
However, it should be noted that the sign of the interference terms in \eqref{eqn:PDCfprmalinterference} changes replacing  $a_0(q)$ with $ia_0(q)$. 
Therefore, we have
	\begin{equation}
		\begin{split}
			&S^{(det)}(q)=S^{(det)}(q;a_0)+S^{(det)}(q;ia_0)=\\
			&=a^{*}(-q,z)a(q,z)=|a_0|^2\left(1+2|V(q,z)|^2\right),
		\end{split}
	\end{equation}
from which follows
	\begin{equation}
		S^{(q)}(q)=|V(q,z)|^2=\frac{S^{(det)}(q)-2|a_0(q)|^2}{4|a_0(q)|^2},
	\end{equation}
thereby arriving at solution in two runs of a standard numerical scheme.
Similarly, we will solve the task for coupled processes in the following section.

Fig.~\ref{fig:fig2} illustrates  $S^{(q)}(q)$  as a function of transverse spatial frequency with different values of the shift angle  $\psi$  and distance $d$  between the crystals. 
Also, for comparison, the dependencies corresponding to the PDC in single crystal, described by solution~\eqref{eqn:standardPDCsolution}, and PDC in half-space, which is obtained by solving~\eqref{eqn:PDChalfcrystal}, are presented. 
From Fig.~\ref{fig:fig2} we infer that the integral efficiency of the PDC conversion, as measured by  $S^{(q)}(q)$, is decreased with increasing the separation  $d$. 
This behavior is due to absence of nonlinear interaction in the inter-crystal space. 
The spectrum profile is also modified  by varying these geometrical parameters  --- 
 shifting the lattice crystals has profound effect on the spectrum profile, which is attributed to interference of the fields originated from different crystals. 
For example, considering two configurations~--- without a shift ($\psi=0$)  and a maximum shift ($\psi=\pi$), 
presented in Fig.~\ref{fig:fig2}a and Fig.~\ref{fig:fig2}d, respectively, we see  the characteristic fringe-like pattern occurred in the shifted structure.

\section{Coupled PDC and SFG processes}\label{sec:coupledprocesses}

We now consider a more complex nonlinear interaction,  comprising the degenerate PDC process, supported by one of the crystals: 
	\begin{equation}\label{eqn:PDCrelations2}
		\omega_p=2\omega_1,
	\end{equation}
which is accompanied by the process of SFG:
	\begin{equation}\label{eqn:SFGrelations}
		\omega_1+\omega_p=\omega_3=3\omega_1,
	\end{equation}
evolving in the other crystal of the binary structure. In this coupled interaction the two-color field of the carrier frequency $\omega_1$  and  the tripled  frequency $\omega_3=3\omega_1$   is produced. 
(The frequency relations between the generated modes can be different in general in the case of two pump waves).
Different QPM conditions for PDC and SFG require that the crystals have different modulation periods.  
Assuming that the lattice of crystal $1$ compensates for the mismatch occurred in PDC: $\Delta{}k_d=k_p-2k_1$, 
while the lattice of crystal $2$ compensates for the mismatch for the SFG:  $\Delta{}k_u=k_p+k_1-k_3$,
the respective modulation periods are calculated as $\Lambda_1=2\pi/|\Delta{}k_d|$  and  $\Lambda_2=2\pi/|\Delta{}k_u|$.

Let the crystals $1$ and $2$ be shifted along axis $z$ in the opposite directions by  $\delta{}z_1$ and  $\delta{}z_2$  in such a way that  $\Delta{}k_d\delta{}z_1/2=-\Delta{}k_u\delta{}z_u/2=\psi$. 
Using \eqref{eqn:2Dnonlinearity} and the QPM conditions for the two nonlinear processes, we obtain the following equations for the field amplitudes $A_1(x,z)$   and $A_3(x,z)$    at frequencies  $\omega_1$  and  $\omega_3$:
	\begin{equation}\label{eqn:coupledamplitudes}
		\left\{
			\begin{split}
				\frac{\partial{}A_1}{\partial{}z}+\frac{i}{2k_1}\frac{\partial^2{}A_1}{\partial{}x^2}&=\beta_d\overline{g}e^{-i\psi}\theta\left(-x-d/2\right)A_pA_1^{*}+\\
				&+\beta_u^{(1)}\overline{g}e^{i\psi}\theta\left(x-d/2\right)A_p^{*}A_3,\\
				\frac{\partial{}A_3}{\partial{}z}+\frac{i}{2k_3}\frac{\partial^2{}A_3}{\partial{}x^2}&=\beta_u^{(2)}\overline{g}e^{i\psi}\theta\left(x-d/2\right)A_pA_1,
			\end{split}
			\right.
	\end{equation}
where  $\beta_d=\omega_1\chi/2nc$  and  $\beta_u^{(m)}=\omega_m\chi/2nc$  are the nonlinear coefficients,  $\chi$ and $n$ is the effective quadratic susceptibility and refractive index of the crystals,  for simplicity, 
assumed identical for the processes involved,   
$k_m=\omega_mn_m/c$  is the wavenumber at frequency   $\omega_m$.
As a result, we have: $\beta_d=\beta_u^{(1)}=\beta$  and  $\beta_u^{(3)}=3\beta$.

As before, the field dynamics is analyzed in the spatial Fourier space. 
Using expansion \eqref{eqn:FourierExpansion}, from \eqref{eqn:coupledamplitudes} we obtain the following set of equations for amplitudes $a_1(q,z)$   and   $a_3(q,z)$:
	\begin{equation}\label{eqn:PDC_SFGfourier}
	\left\{
		\begin{split}
			\frac{\partial{}a_1(q,z)}{\partial{}z}-i\varepsilon_1a_1(q,z)=&\\
			=\frac{1}{2}\sigma\overline{g}\left[e^{i(\varphi_p-\psi)}a_1^{*}(-q,z)+e^{-i(\varphi_p-\psi)}a_3(q,z)-\right.&\\
			-ie^{i(\varphi_p-\psi)}e^{-iqd/2}H\left(e^{iqd/2}a_1^{*}(-q,z)\right)+&\\
			\left.+ie^{-i(\varphi_p-\psi)}e^{iqd/2}H\left(e^{-iqd/2}a_3(q,z)\right)\right],&\\
			\frac{\partial{}a_3(q,z)}{\partial{}z}-i\varepsilon_3a_3(q,z)=&\\
			=\frac{3}{2}\sigma\overline{g}e^{i(\varphi_p+\psi)}\left[a_1(q,z)+ie^{iqd/2}H\left(e^{-iqd/2}a_1(q,z)\right)\right],&
		\end{split}
		\right.
	\end{equation}
where  $\varepsilon_1=q^2/2k_2$,  $\varepsilon_3=q^2/2k_3=\varepsilon_1/3$   with the rest of the parameters having the same meaning as in \eqref{eqn:fourierPDC}.
\eqref{eqn:PDC_SFGfourier} are not amenable to analytical solution and we solve them numerically using the approach  presented in the previous section. 

Before proceeding further, for comparison reason, let us introduce aperiodic nonlinear crystals, in which the coupled processes under consideration can be implemented. Due to aperiodic modulation of the nonlinearity sign in the longitudinal direction distinct multiple peaks in the lattice spectrum can occur in the aperiodic structures. We will compare the efficacy of the binary structures with the crystals having aperiodic modulation.

Several design methods that enable structures supporting multiple nonlinear processes simultaneously have been proposed. The method of superposition of modulation of nonlinearity sign (SMNS) has been proved to be a versatile approach to construct aperiodic crystals~\cite{Chirkin,ShutovChirkin,ChirkinShutov2}. 
According to this method, to compensate for multiple $N$ phase mismatches by the structure  $\Delta{}k_j$ ($j=\overline{1,N}$), the modulating function is calculated~\cite{ChirkinShutov2}:
	\begin{equation}
		g_{AS}(z)=\text{sign}\left[\sum_{j=1}^NC_j\cos\left(\frac{2\pi}{\Lambda_j}+\varphi_j\right)\right]
	\end{equation}
where $\Lambda_j=2\pi/|\Delta{}k_j|$  is the lattice period that compensates the mismatch $\Delta{}k_j$, $\varphi_j$  is the phase of the partial harmonic in the superposition,  $C_j$ is the amplitude that defines the contribution of the harmonic to the lattice, subscript $j$ marks the mismatch $\Delta{}k_j$  ($j=\overline{1,N}$). 
By defining parameters  $C_j$ and $\varphi_j$  at the construction stage, the coupling coefficients and their phases can be judiciously engineered.

Following the SMNS method, calculation of the peak amplitudes that enter the nonlinear equations (\eqref{eqn:2Dnonlinearity}) gives the following result~\cite{ChirkinShutov2}:
	\begin{equation}\label{eqn:aperiodicamplitudes}
		\overline{g}_{AS1}=\frac{2}{\pi}I(C_1,C_2),\qquad\overline{g}_{AS3}=\frac{2}{\pi}I(C_2,C_1),
	\end{equation}
where  $I(C_1,C_2)=\int_0^{+\infty}J_1(C_1x)J_0(C_2x)/x{}dx=2(E(r^2)+K(r^2)(r^2-1))/\pi{}r$, $r=C_2/C_1$,  $K(x)$ and $E(x)$ are the elliptical integral of the first and second kinds, respectively. 
Since the crystals in the binary structure have equal lattice peak amplitudes ($\overline{g}=2/\pi$), 
in \eqref{eqn:aperiodicamplitudes} we set $C_1=C_2$   for the aperiodic structure has equal peaks as well. Using \eqref{eqn:aperiodicamplitudes}, the lattice amplitudes that compensate for mismatches  $\Delta{}k_d$ and $\Delta{}k_u$  are obtained:   $\overline{g}_{AS1}=\overline{g}_{AS3}=(2/\pi)^2<\overline{g}$, so that
the peak amplitudes of an aperiodic lattice is lesser than the ones of a periodic one.

Therefore, for the case of aperiodic structure, we have the following set of equations that describe the coupled processes~\cite{Makeev}: 
	\begin{equation}\label{eqn:aperiodiceqs}
		\left\{
			\begin{split}
				\frac{\partial{}a_1(q,z)}{\partial{}z}-i\varepsilon_1a_1(q,z)=&\\
				\sigma\overline{g}_{AS1}a_1^{*}(-q,z)+\sigma\overline{g}_{AS3}a_2(q,z),&\\
				\frac{\partial{}a_3(q,z)}{\partial{}z}-i\varepsilon_3a_3(q,z)=3\sigma\overline{g}_{AS3}a_1(q,z).&
			\end{split}
		\right.
	\end{equation}
\eqref{eqn:aperiodiceqs} are rather simple and has an analytical solution (see, for example,~\cite{Makeev}). In the notations adopted above, the form of the solution is the following:
	\begin{equation}\label{eqn:coupledForm}
		\begin{split}
			a_1(q,z)=U_{11}(q,z)a_{10}(q)+V_{11}(q,z)a_{10}^{*}(-q)&\\
			+U_{13}(q,z)a_{30}(q)+V_{13}(q,z)a_{30}^{*}(-q),&\\
			a_3(q,z)=U_{31}(q,z)a_{10}(q)+V_{31}(q,z)a_{10}^{*}(-q)&\\
			+U_{33}(q,z)a_{30}(q)+V_{33}(q,z)a_{30}^{*}(-q),&
		\end{split}
	\end{equation}
where the transfer functions $U_{mn}(q,z)$  and $V_{mn}(q,z)$  are mutually related by
	\begin{equation}\label{eqn:coupledCommutation}
			\begin{split}
				|U_{m1}(q,z)|^2+|U_{m3}(q,z)|^2&\\
				-|V_{m1}(q,z)|^2-|V_{m3}(q,z)|^2=1,&
			\end{split}
	\end{equation}
($m=1,3$), which follow from commutation relations, associated with the Fourier amplitudes; the field amplitudes  $a_1(q,z)$, $a_3(q,z)$, $a_1^{*}(q,z)$   and $a_3^{*}(q,z)$  correspond to photon creation and annihilation operators. 
Solution of the  form of solution \eqref{eqn:coupledForm} holds true for the case of binary structure.

In the case of  vacuum input, the spatial correlation \eqref{eqn:PDCcorrelator} is applicable to both the low- and high-frequency modes: $\langle{}a_{m0}^{*}(-q')a_{n0}(q'')\rangle=S_m(q')\delta_{mn}\delta(q'+q'')$    ($m=1,3$), where  $\delta_{mn}$ is the Kronecker symbol appeared since vacuum modes of different frequencies are uncorrelated.

From the quantum point of view, the correct spectral densities $S_m(q)$  are derived using \eqref{eqn:coupledForm} using operator mathematics and the vacuum state condition   
$\langle{}a_{m0}^{\dagger}(-q)a_{m0}(q)\rangle=0$:
	\begin{equation}\label{eqn:quantumS}
			S_m^{(q)}(q)=|V_{m1}(q,L)|^2+|V_{m3}(q,L)|^2,\quad(m=1,3).
	\end{equation}

In the classical domain with random noise model for the vacuum amplitudes, using \eqref{eqn:coupledForm} and \eqref{eqn:coupledCommutation}, 
the spatial spectral densities read:
	\begin{equation}\label{eqn:classicalS}
		\begin{split}
			S_m^{(c)}(q)=\langle\hat{a}_m^{*}(-q)\hat{a}_m(q)\rangle=&\\
			\left[1+2(|V_{m1}(q,L)|^2+|V_{m3}(q,L)|^2)\right]|a_0|^2,&
		\end{split}
	\end{equation}
($m=1,3$).

To relate the solution of \eqref{eqn:PDC_SFGfourier} obtained by numerical simulation with the correct solution, dictated by quantum physics, we apply the same approach, that has been used for single PDC process. 
It is evident from \eqref{eqn:PDC_SFGfourier} that the quadratic combinations of field amplitudes, when considered as classical values, include interference terms proportional to         
$a_{10}^2$, $a_{30}^2$, $a_{10}a_{30}$, $a_{10}^{*}a_{30}$ and the complex conjugates.
To eliminate them, we solve the equation numerically $4$ times with the following set of initial values: 1)  $a_{10}=a_{30}=a_0$, 2) $a_{10}=-a_{30}=a_0$,  3)  $a_{10}=a_{30}=ia_0$  and 4) $a_{10}=-a_{30}=ia_0$. 
As a result of this, we obtain $4$ dependencies: 
$S^{(det)}_m(q;a_0,a_0)$,  $S^{(det)}_m(q;a_0,-a_0)$, $S^{(det)}_m(q;ia_0,ia_0)$,  $S^{(det)}_m(q;ia_0,-ia_0)$,  which total, $S^{(det)}_m(q)$, successfully eliminates the unwanted terms 
giving quadrupled classical spectral densities \eqref{eqn:classicalS}.
Therefore, from \eqref{eqn:quantumS} and \eqref{eqn:classicalS} readily follows:
	\begin{equation}
		S_m^{(q)}(q)=\frac{S_m^{(det)}(q)-4|a_0|^2}{8|a_0|^2},\quad(m=1,3)
	\end{equation}
where $S_m^{(det)}(q)=S^{(det)}_m(q;a_0,a_0)+S^{(det)}_m(q;a_0,-a_0)+S^{(det)}_m(q;ia_0,ia_0)+S^{(det)}_m(q;ia_0,-ia_0)$.

\begin{figure}[htbp]
\centering
\includegraphics[width=0.65\linewidth]{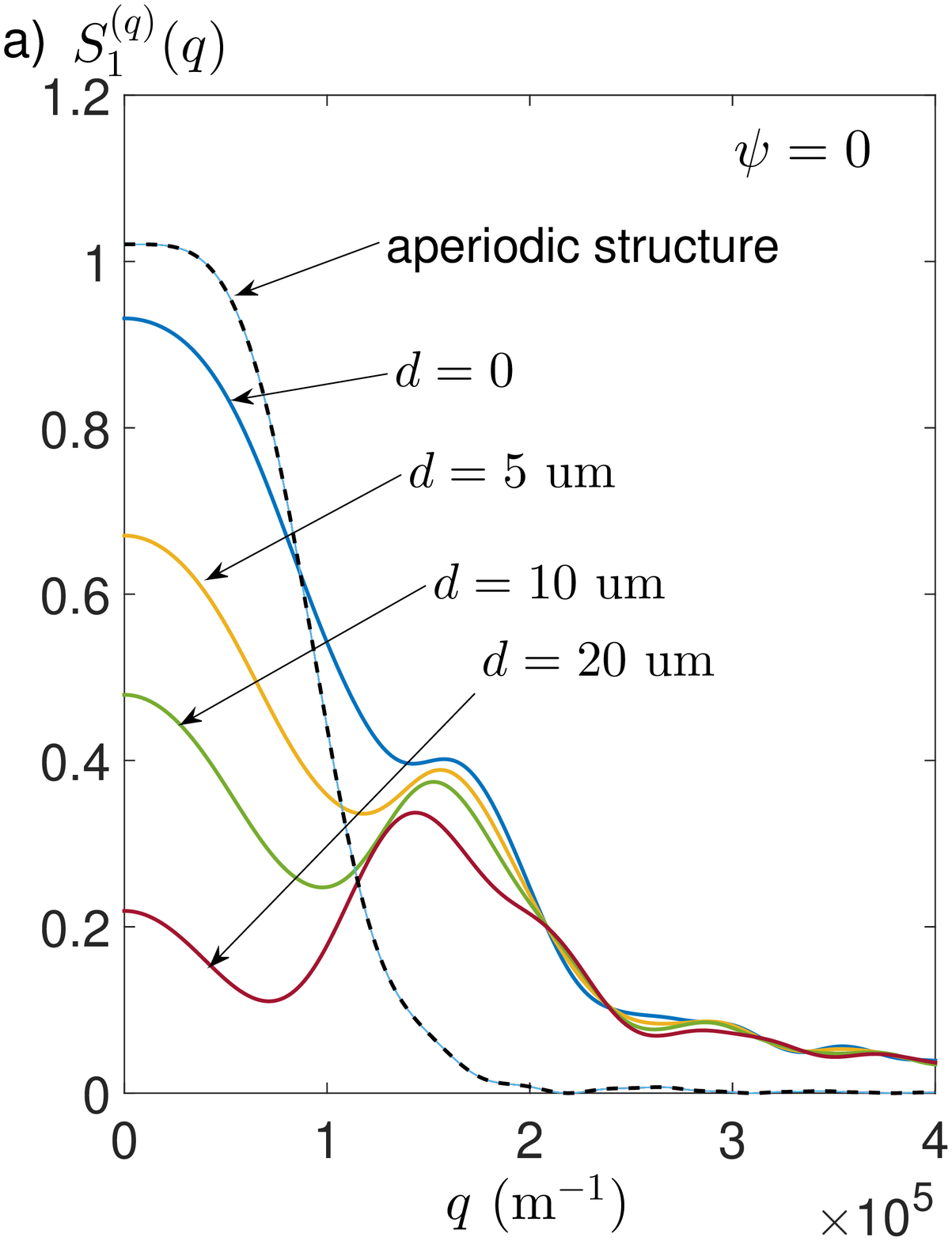}\\
\includegraphics[width=0.65\linewidth]{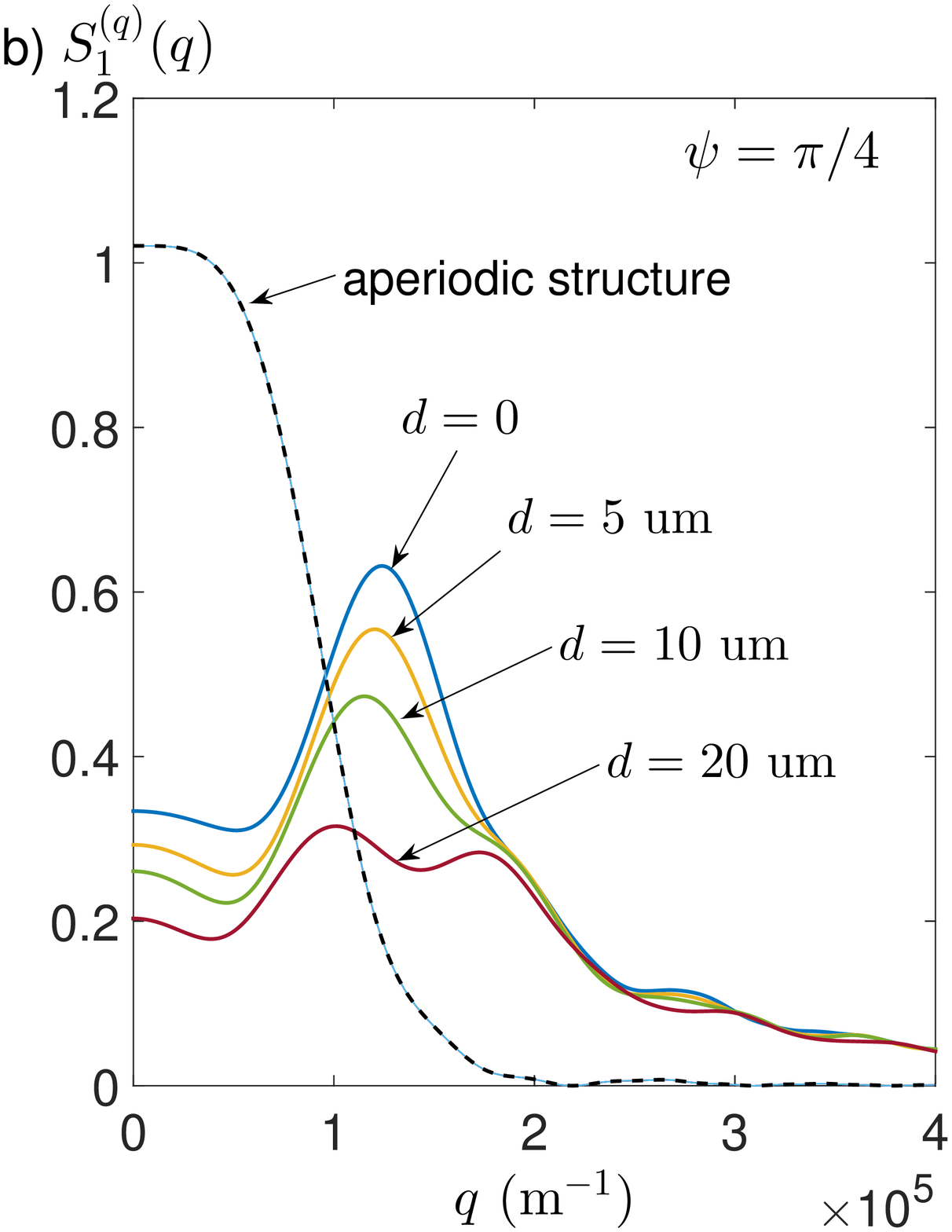}\\
\includegraphics[width=0.65\linewidth]{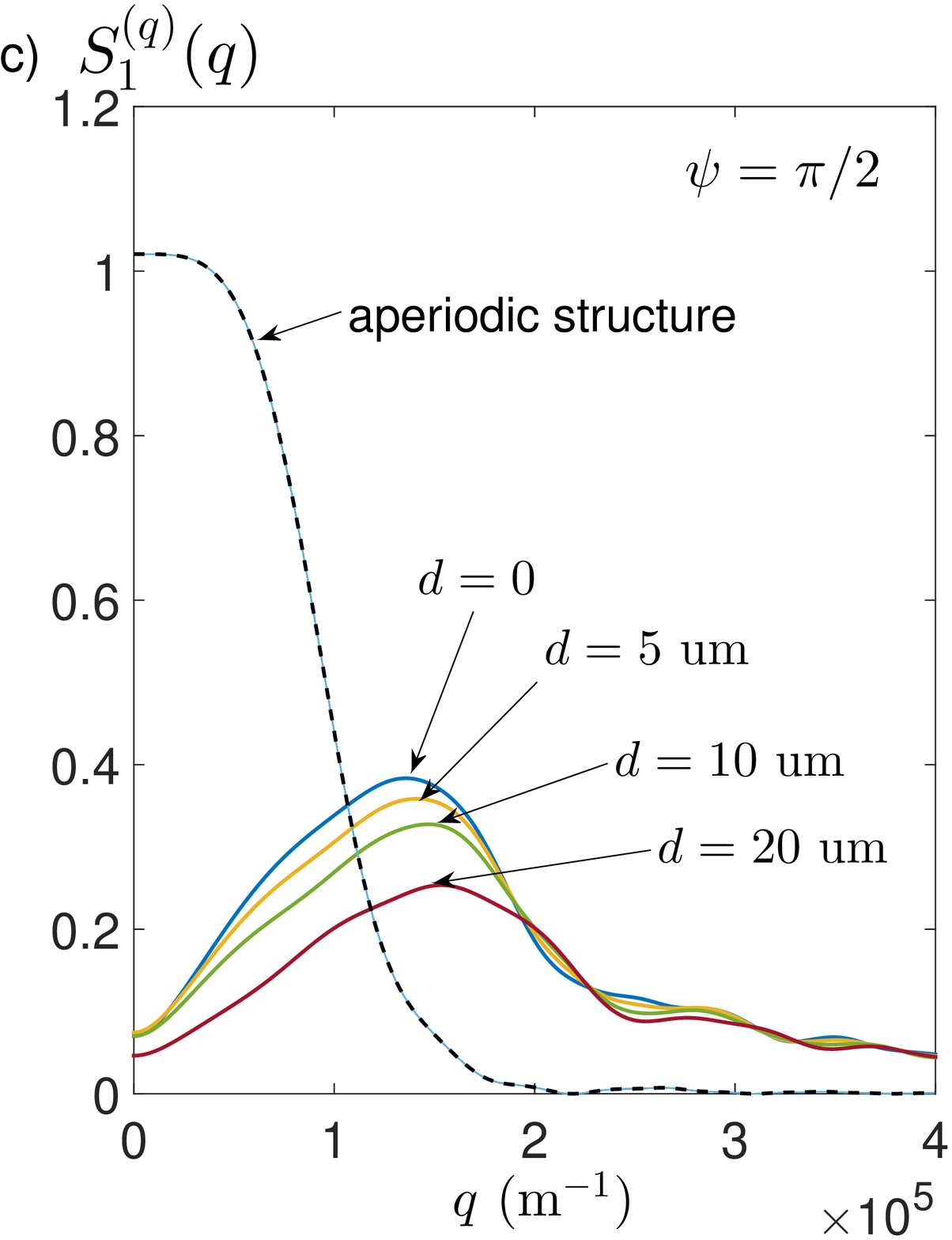}
\caption{Spectral density $S_1^{(q)}(q)$ as a function of the spatial frequency at different crystal shift values: 
a) $\psi=0$, b) $\psi=\pi/4$, c) $\psi=\pi/2$.
The dependencies drawn in dashed correspond to the field generated in 
the coupled interactions implemented in the aperiodic structure described by~\eqref{eqn:aperiodiceqs} 
with lattice spectrum amplitude  $\overline{g}_{AS1}=\overline{g}_{AS3}=(2/\pi)^2$. 
The pump phase  $\varphi_p=\psi$. The rest of the parameters are the same as in Fig.~\ref{fig:fig2}.
}
\label{fig:fig3}
\end{figure}

\begin{figure}[htbp]
\centering
\includegraphics[width=0.65\linewidth]{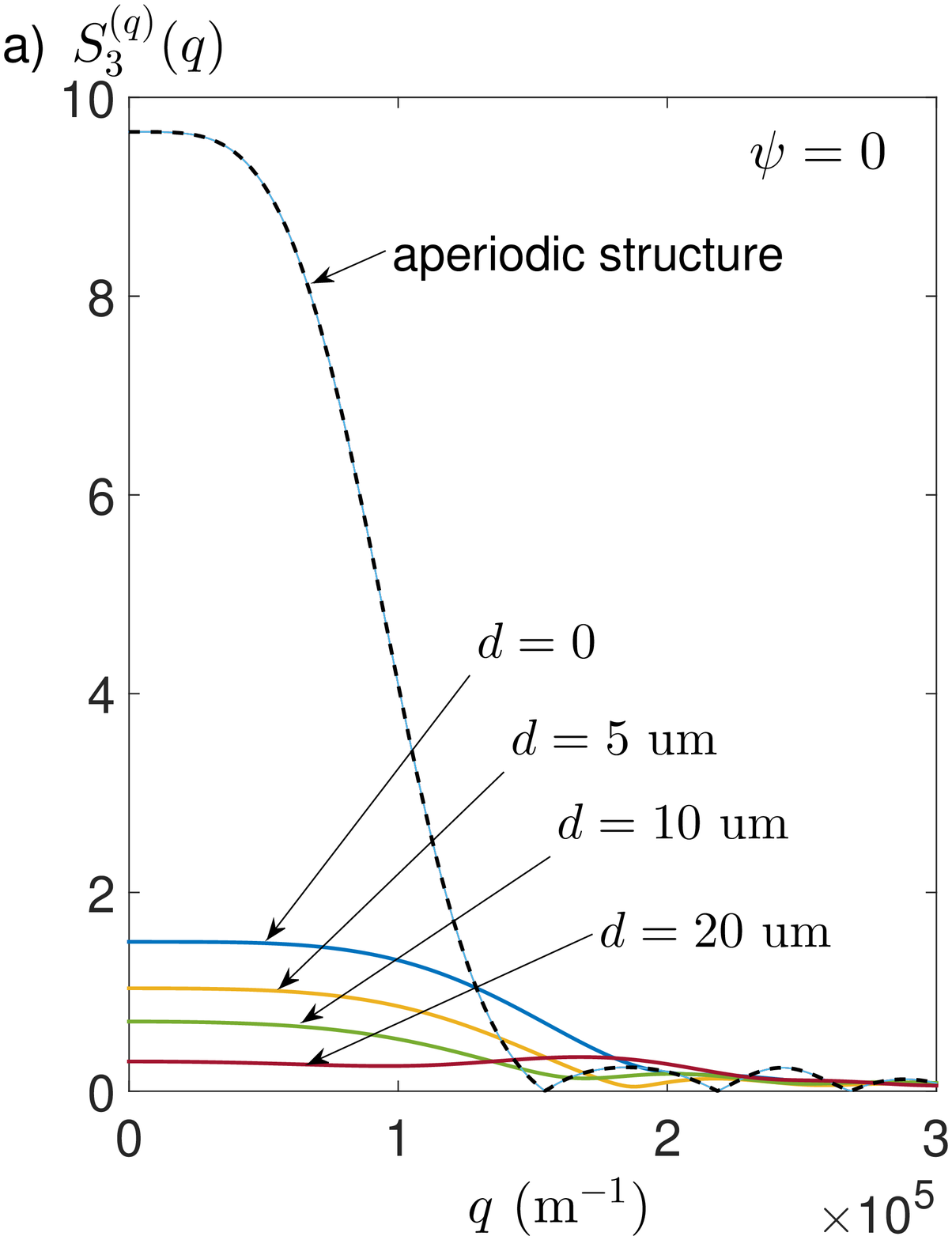}\\
\includegraphics[width=0.65\linewidth]{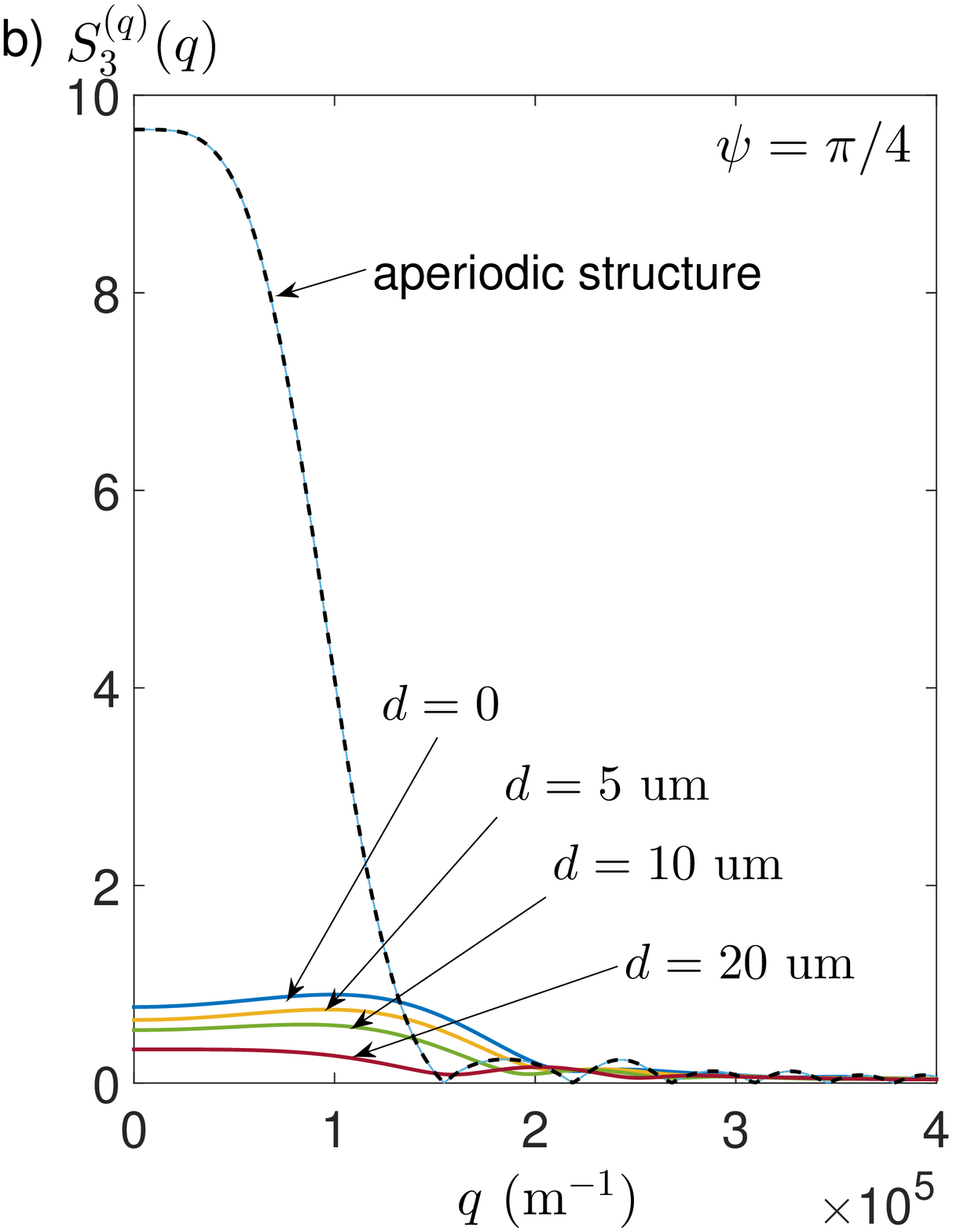}\\
\includegraphics[width=0.65\linewidth]{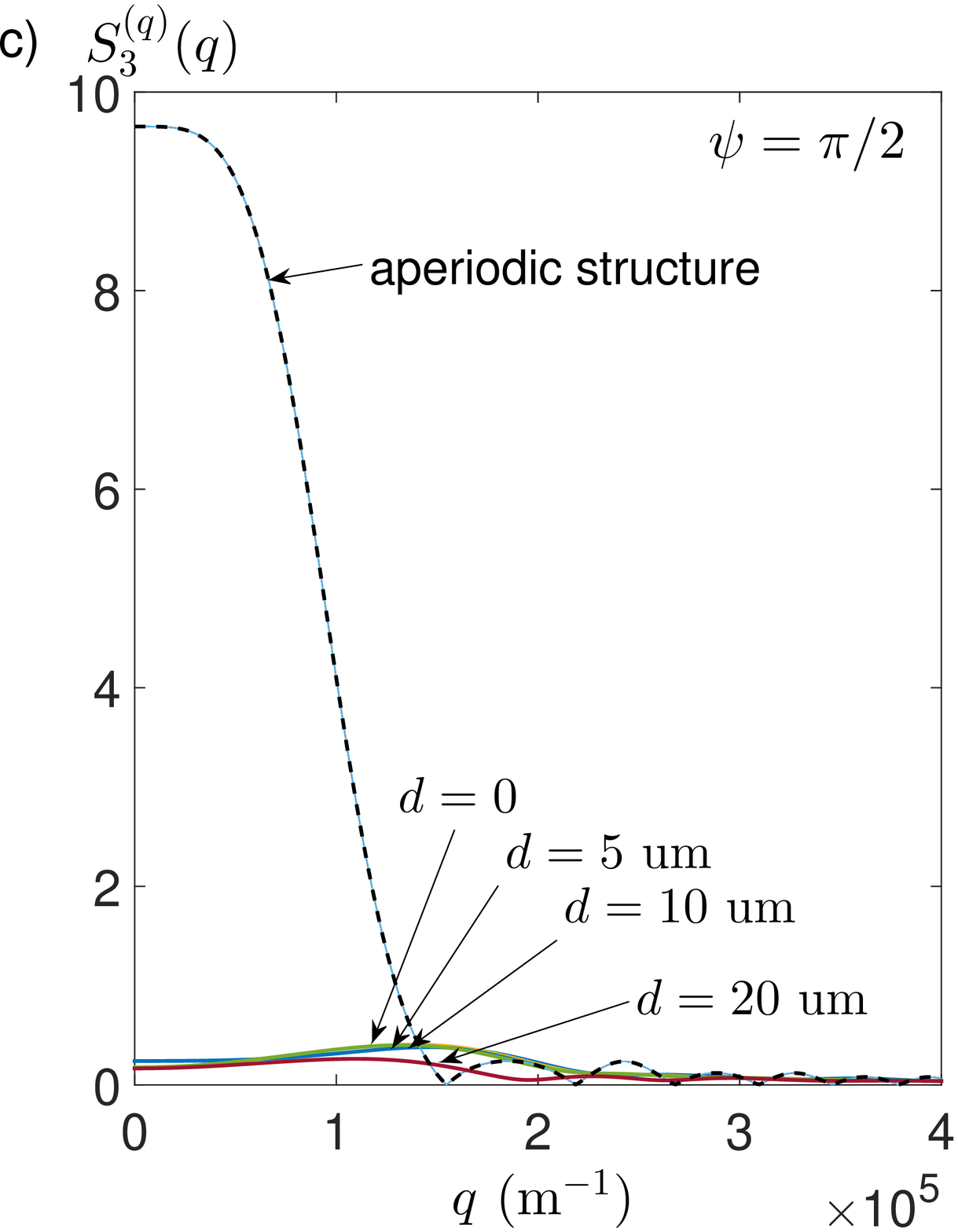}
\caption{Spectral density $S_3^{(q)}(q)$ as a function of the spatial frequency at different crystal shift values: 
a) $\psi=0$, b) $\psi=\pi/4$, c) $\psi=\pi/2$.
The relevant parameters are the same as in Fig.~\ref{fig:fig3}}
\label{fig:fig4}
\end{figure}

The spatial spectral densities $S_1^{(q)}(q)$ and  $S_3^{(q)}(q)$ are plotted in Fig.~\ref{fig:fig3} and Fig.~\ref{fig:fig4}, respectively,
at different values of the crystal shift  and the distance between the crystals. 
Comparing the dependencies in the figures with the ones corresponding to the fields generated in aperiodic structures, described by~\eqref{eqn:aperiodiceqs}, 
it is evident that the integral efficiency of field formation in the binary structure is always lower than in the aperiodic structure.
Especially, we notice that in the aperiodic structure with the lattice parameters under study, 
the high-frequency field  is brighter than the low-frequency field, 
which is due to the higher value of the coupling strength responsible for SFG.
In the case of the binary structure,  the same behaviour holds true, however,
the difference in efficiency, as quantified by the ratio between the corresponding spectral maxima, is not so high.
Taking into account that in \eqref{eqn:PDC_SFGfourier} and \eqref{eqn:aperiodiceqs}  the parameters responsible for diffraction obey $\varepsilon_3=\varepsilon_1/3$, 
the  diffraction effects  in the high-frequency range are  manifested to lesser degree.
Therefore, the interference fringe-like behaviour observed  in the field spectral density and produced by single PDC in the binary structure 
is not so apparent in the case of the coupled PDC and SFG (compare Fig.~\ref{fig:fig2}c) and Fig.~\ref{fig:fig3}c)).
Moreover, the corresponding spatial density of high-frequency modes ( Fig.~\ref{fig:fig4}c)) does  not manifest interference behaviour at all.

\section{Conclusion}\label{sec:conclusion}

To summarise, we have proposed the binary nonlinear structure as means for the implementation of multiwave optical processes. 
Our study suggests that despite each crystal in the structure supports distinct set of nonlinear processes, coupling between modes originated from the two crystals can be attained near the inter-crystal interface, due to loosely constraints of phase-matching. 
Also, through consideration of the coupled PDC and SFG processes, we have shown that the field generation and conversion efficiencies in the binary structure is somewhat lower than in the corresponding nonlinear crystals without transverse inhomogeneity.

Other configurations of the binary structures, which are different in constituent materials and modulation functions, can also be suggested. 
In particular, the pulsed regime of parametric interaction with chirped crystals~\cite{Suchowski} can be of interest, due to the intertwining of the material dispersion of the crystals, that can alter the pulse dynamics. Also, the capability of the binary structures to host structured film materials, 
for example, metamaterials~\cite{Kivshar}, in space between the crystals makes them attractive for development of potentially new optical sources with unique temporal and spatial characteristics.

\section{Acknowledgements}\label{sec:acknowledgements}

We acknowledge partial support from the Russian Foundation for Basic Research under the grant No 14-02-00458.

\end{document}